\documentclass[aip,pop,preprint]{revtex4-1}
\usepackage{amsmath}
\usepackage{amssymb}
\usepackage{upgreek}
\usepackage{graphicx}
\usepackage{bm}

\begin{document}

\title{Effect of laser polarization on QED cascading}
\author{V.~F.~Bashmakov}
\affiliation{Institute of Applied Physics, Russian Academy of Sciences, 
603950 Nizhny Novgorod, Russia}
\affiliation{University of Nizhny Novgorod, 23 Gagarin Avenue, Nizhny
Novgorod 603950, Russia}
\author{E.~N.~Nerush}
\affiliation{Institute of Applied Physics, Russian Academy of Sciences, 
603950 Nizhny Novgorod, Russia}
\affiliation{University of Nizhny Novgorod, 23 Gagarin Avenue, Nizhny
Novgorod 603950, Russia}
\author{I.~Yu.~Kostyukov}
\email{kost@appl.sci-nnov.ru}
\affiliation{Institute of Applied Physics, Russian Academy of Sciences, 
603950 Nizhny Novgorod, Russia}
\affiliation{University of Nizhny Novgorod, 23 Gagarin Avenue, Nizhny
Novgorod 603950, Russia}
\author{A.~M.~Fedotov}
\affiliation{National Research Nuclear University MEPhI, Moscow, 115409, 
Russia}
\author{N.~B.~Narozhny,}
\affiliation{National Research Nuclear University MEPhI, Moscow, 115409, 
Russia}

\begin{abstract}
Development of QED cascades in a standing electromagnetic wave for circular and linear polarizations is simulated numerically with a 3D PIC-MC code. It is demonstrated that for the same laser energy the number of particles produced in a circularly polarized field is greater than in a linearly polarized field, though the acquiring mean energy per particle is larger in the latter case.
The qualitative model of laser-assisted QED cascades is extended by including the effect
of polarization of the field. It turns out that cascade dynamics is notably more complicated in the case
of linearly polarized field, where separation into the qualitatively different "electric" and "magnetic" regions
(where the electric field is stronger than the magnetic field and vice versa) becomes essential. In the "electric" regions acceleration is suppressed and moreover the high-energy electrons are even getting cooled by photon emission. The volumes of the "electric" and "magnetic" regions evolve periodically in time, and so does the cascade growth rate. In contrast to the linear polarization the charged particles can be accelerated by circularly polarized wave even in "magnetic region". The "electric" and "magnetic" regions do not evolve in time and cascade growth rate almost does not depend on time for circular polarization.

\end{abstract}


  \maketitle

\section{Introduction}

Quantum electrodynamical  (or electromagnetic) cascades
play an important role in astrophysical phenomena. Cascades 
initiated by high energy cosmic rays produce electromagnetic 
showers in magnetospheres and atmospheres of planets \cite{Rossi1953}. 
It is generally believed that cascading is a key mechanism 
of electron-positron plasma production at the neutron stars 
\cite{Daugherty1982}. Recently QED cascading in strong laser 
field has attracted significant attention \cite{Bell2008,Fedotov2010,PRL2011}.
Interest to laser-assisted QED cascading comes due 
to a rapid progress in laser technology which opens 
opportunities to study the high-field QED effects under the 
laboratory conditions with the upcoming high-power laser facilities \cite{ELI,XCELS}.

A cascade develops as a sequence of elementary QED processes:
photon emission by relativistic charged particles in the field of
a nucleus or in an external electromagnetic field alternates with photon
decay by a pair production. Such an order of the events leads to an 
avalanche-like production of electron-positron plasma and $\gamma$-quanta.
In the case of electromagnetic showers the energy of the cascade particles
is retrieved exclusively from the energy of the incoming cosmic ray. However, the cascade
energy can also be gained  from the external electromagnetic field
as, e.g. in the vicinity of a surface of a pulsar or in 
laser-assisted QED cascades. In the latter case the electrons and the positrons
produced during cascade development are accelerated in the laser field. 

Laser acceleration is capable for boosting up the energy of 
the charged particles and, more notably, for turning them 
around transversely to the field, thus increasing dramatically 
the probabilities of QED processes and, accordingly, 
the cascading rate. If the plasma resulting from cascading 
becomes rather dense, the self-generated plasma field can 
become even as strong as the laser field itself. In such a case 
the laser field can be significantly depleted because of
the avalanche-like electron-positron plasma production and
$\gamma$-ray emission \cite{PRL2011}. In this way QED cascades
may limit the attainable intensity of the focused laser pulses 
\cite{Fedotov2010}.

The QED cascade can be seeded either by external particles 
injected in the laser focal spot, or even by the pairs created 
due to vacuum breakdown. Electron-positron plasma can be produced 
also directly via vacuum breakdown in the strong electromagnetic 
field, but in order to produce dense enough electron-positron 
plasma by this way the field strength
has to be of the order of the QED critical field, 
$E_{cr}=m^{2}c^{3}/e\hbar\simeq1.3\times10^{16}$ V/cm
\cite{Sauter1931,Narozhny2004,Bulanov2006}, where $e>0$ and $m$ are 
the value of electron charge and the electron mass, respectively. 
However, the cascades in the presence of a seed appear already 
at much lower values of the field strength.

One of the key QED parameters that determine the probability 
of photon emission and radiation regime is
and radiation regime is \cite{Nikishov,Landau4}
\begin{eqnarray}
\chi & = & \frac{e\hbar}{m^{3}c^{4}}\sqrt{\left(\frac{\varepsilon
\mathbf{E}}{c}+\mathbf{p}\times\mathbf{B}\right)^{2}-\left(\mathbf{p}
\cdot\mathbf{E}\right)^{2}},
\label{chi}\end{eqnarray}
  where $\mathbf{E}$ and $\mathbf{B}$ are the electric and the magnetic
fields, $\varepsilon$ and $\mathbf{p}$ are the energy and the momentum
of an electron (positron). As was discussed in 
\cite{Fedotov2010,Narozhny2004,Bulanov2006}, below the QED critical field and 
for optical range it is enough to use the locally constant field 
approximation. Then the probability of photon emission by
an electron (positron) with energy $\varepsilon$ is readily given by the
formula \cite{Baier}
\begin{eqnarray}
W_{rad} & = & \frac{\alpha m^{2}c^{4}}{3^{3/2}\pi\hbar\epsilon}
\int_{0}^{\infty}du\frac{5u^{2}+7u+5}{(1+u)^{3}}K_{2/3}\left(\frac{2u}
{3\chi}\right),\label{W1}\\
W_{rad} & \approx & 1.44\frac{\alpha m^{2}c^{4}} {\pi\hbar\varepsilon}\chi,\;\chi\ll1,\label{W2}\\
W_{rad} & \approx & 1.46\frac{\alpha m^{2}c^{4}}{\hbar\varepsilon}
\chi^{2/3},\;\chi\gg1,
\label{W3}\end{eqnarray}
  where $\alpha=e^2/\hbar c$ is the fine structure constant and 
$K_{\nu}(x)$ is the McDonald function \cite{Abramowitz}.
The radiation process can be treated classically in the
limit $\chi\ll1$. In this limit the photon emission probability is
determined by Eq.~(\ref{W2}). The quantum nature of photon emission
manifests itself (for example, through the spin and the recoil effects) 
at high intensities or for high energy,  $\chi\geq1$. In the limit $\chi\gg1$ the
probability becomes a nonlinear function of the electron energy and the
electromagnetic field strength and is reduced to Eq.~(\ref{W3}).

Pair photoproduction in a strong electromagnetic field is a cross 
channel of photon emission \cite{Landau4}. It's determined by similar 
QED parameter $\chi_{ph}$, which is defined by Eq.~(\ref{chi}), 
where $\varepsilon$ and $\mathbf{p}$ are substituted by 
the photon energy $\varepsilon_{ph}$ and the photon 
momentum $\mathbf{p}_{ph}$. The probability of pair production is given 
by the formulas (see also \cite{Nikishov})
\begin{eqnarray}
W_{pair} & = & \frac{\alpha m^{2}c^{4}}{3^{3/2} \pi\hbar\varepsilon_{ph}}
\int_{0}^{1}du\frac{9-u^{2}}{1-u^{2}}K_{2/3}\left(\frac{8u/3 \chi_{ph} }{
1-u^{2}}\right),\label{Wg1}\\
W_{pair} & \approx & 0.23\frac{\alpha m^{2}c^{4}}{\hbar\varepsilon_{ph}}
\chi_{ph}\exp\left(-\frac{8}{3\chi_{ph}}\right),\;\chi_{ph}\ll1,\label{Wg2}\\
W_{pair} & \approx & 0.38\frac{\alpha m^{2}c^{4}}{\hbar\varepsilon_{ph}}
\chi_{ph}^{2/3},\;\chi\gg1
\label{Wg3}
\end{eqnarray}

Unlike the photon emission, the pair production probability turns exponentially
small in the quasiclassical limit $\chi_{ph}\ll1$.

Photon emission and pair photoproduction are not efficient if the initial particle
and the electromagnetic wave propagate in the same directions. QED
cascading may occur in a single plane electromagnetic wave if the seed 
counter-propagates the wave, however it decays quickly since the produced electron-positron pairs are pushed by the field mostly along the direction of propagation of the wave. However, it was shown
\cite{Bell2008,Kirk2009} that cascades can develop efficiently in a standing 
electromagnetic wave, which can be generated by two counter-propagating laser pulses.
As the pair production probability vanishes exponentially as the field 
strength decreases, there must exist a vague threshold value of the laser intensity
required for cascading. Estimations show \cite{Fedotov2010} that cascade
development becomes possible for lasers with intensities of the order
of $10^{25}$ W/cm$^{2}$. Numerical simulations \cite{PRL2011} had demonstrated 
that the actual threshold is even lower.

Cascade origination and development is a rather complex phenomenon due to interplay
between the QED and plasma effects, hence in most cases numerical simulations are the only tool to explore it. A typical 
numerical scheme taking proper account of QED effects for modeling laser plasma dynamics combines a 
particle-in-cell (PIC) and Monte-Carlo (MC) methods
 \cite{PRL2011,Timokhin2010,Sokolov2010,Ridgers2012,NerushNIMA2011}. The 
trajectories of the particles and the distribution of the laser-plasma
fields are calculated by the PIC method while the photon emission and
pair photoproduction is modelled with MC method. The validity of PIC and MC
methods is justified because the formation lengths of the processes of emission of $\gamma$-quanta
and pair production are much less than both the laser wavelength and the
mean free path of the cascade particles \cite{Elkina2011,Nerush2011}.
The $\gamma$-quanta can be treated in simulation as particles while 
the low-frequency laser and plasma fields can 
be calculated by integrating Maxwell equations.

PIC-MC simulations have been used to evaluate the laser intensity threshold
for cascade production \cite{Duclous2011,Elkina2011,Nerush2011}, as well as to study the nonlinear
stage of QED cascade with strong plasma absorption of the laser field \cite{PRL2011}.
Up to now, the self-consistent numerical modeling of QED used to be restricted by two dimensions. Obviously, an extension to 3D would be a goal because particle motion in those field configurations which are interesting for applications is usually essentially three-dimensional. Here we report the results of 3D simulation of QED cascading in the field of long counter-propagating laser pulses. In order to study polarization effect and to exclude 
influence of the other laser parameters (like pulse duration and pulse radius) we consider the field configuration, which is close to the standing electromagnetic field. 

The impact of polarization of the laser field on QED cascading 
has been studied with numerical model based on the assumption that electron
radiation losses occur continuously and thus can be governed by some modified version of the Landau-Lifshitz equation \cite{Kirk2009}. Another drawback of of such a model was that it allowed to analyse only   the first generation of cascade particles.
Here we employ 3D PIC-MC simulation, which is free from all such assumptions 
and limitations. We also extended the qualitative model of QED cascade.
In the current paper, we only focus on the early stages of cascade development, 
when the electromagnetic self-field of the arising electron-positron plasma can still be neglected.

The paper is organized as follows. In Sec. II the results of numerical
simulations by 3D PIC-MC code are presented. In order to comment on and explain them, an analytical model is developed in Sec. III. Finally, conclusion and discussion of the results are
collected in Sec. IV.

\section{Numerical Simulation}
\subsection{Circular polarization}

Consider first the QED cascading driven by circularly
polarized (CP) laser pulses. We approximate the wave field by the field
of two counter-propagating long laser pulses, assume that pulses counter-propagate along 
$x$ axis 
and choose the initial condition for the laser field at $t=0$ in the form
\begin{eqnarray}
E_y & = &g(y,z)\left[f_2(x + x_0)-f_2(x-x_0)\right],\label{ecp1} \\
E_z &=&g(y,z)\left[- f_1(x+x_0)-f_1(x-x_0)\right],\label{ecp11}\\
B_y &=&g(y,z)\left[-f_1(x - x_0)+f_1(x+x_0)\right],\label{bcp1} \\
B_z &=&g(y,z)\left[f_2(x + x_0)+f_2(x-x_0)\right],\label{bcp11}
\\
g\left( y,z \right) &=& a_{0}\cos^2\left(\frac{y}{\sigma_{r}} \right)
\cos^2\left(\frac{z}{\sigma_{r}} \right)
\\
f_1 \left( x \right) &=& \cos \left( x \right) \cos^{2}\left(\frac{x} 
{\sigma_{x}}\right)
\\
f_2 \left( x \right) &=& \sin \left( x \right) \cos^{2}\left(\frac{x} 
{\sigma_{x}}\right)
\end{eqnarray}
  where the field strengths are normalized to $mc\omega_{L}/|e|$,
$a_{0}=eE_{0}/mc\omega_{L}$, $E_{0}$ is the electric field amplitude
of a single laser pulse, $\omega_{L}$ is the laser pulse cyclic frequency,
$x_{0}$ is a half of the initial distance between the laser pulses. The parameters
of simulations are $\sigma_{x}=53\lambda$, $\sigma_{r}=3\lambda$,
$x_{0}=5\lambda$, where $\lambda=2\pi c/\omega_{L}=0.91 \mu \text{m}$
is the laser wavelength. The cascade is initiated by a bunch of MeV-photons
moving along the $x$ axis with a center located initially at $y=z=0$ and $x=-x_{0}$. The length  and the radius of the photon bunch are $4\lambda$ and
 $0.1\lambda$, respectively. Cascading is explored for two values of laser intensity, corresponding to $a_0=2.0\times 10^3$ and $a_0=2^{-1/2}\cdot8000=5.66\times 10^3$. 
 
\begin{figure}
\includegraphics[width=8cm]{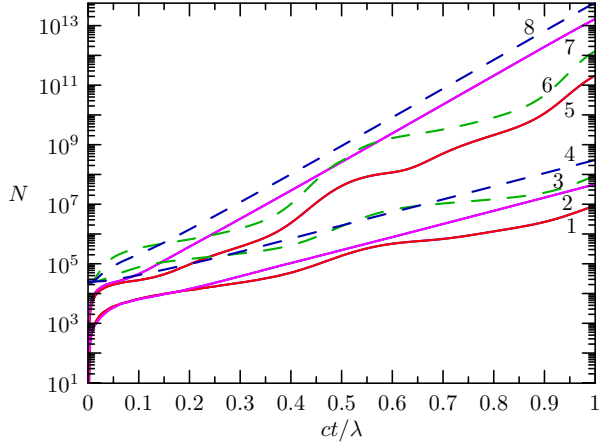}
\caption{Time dependence of the number of electrons (positrons)
in a QED cascade in LP laser field for 
$a_{0}=2.83\times10^3$
(red solid line 1), and $a_{0}=8.0\times10^3$ (red solid line 5) and
in CP laser field for $a_{0}=2.0\times10^3$
(magenta solid line 2) and $a_{0}=5.66\times10^3$ (red solid line 7).
Time dependence of the number of $\gamma$-quanta 
in a QED cascade in LP laser field for 
$a_{0}=2.83\times10^3$
(green dashed line 3), for $a_{0}=8.0\times10^3$ (green dashed line 6) and
in CP laser field for $a_{0}=2.0\times10^3$ (blue dashed line 4),
and $a_{0}=5.66\times10^3$ (blue dashed line 8).}
\label{fig-N}
\end{figure}

\begin{figure}
\includegraphics[width=8cm]{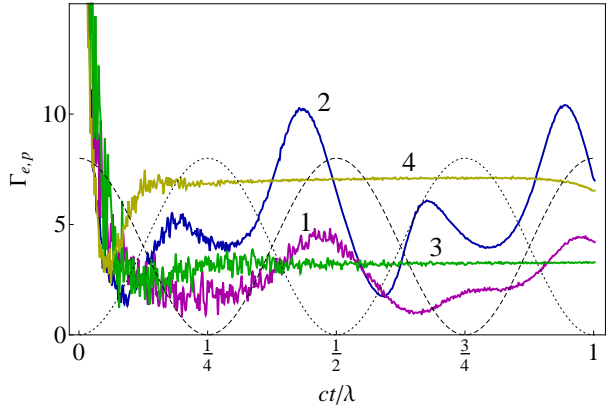}
\caption{Evolution of the growth rates of pairs (solid lines)  in LP laser field with $a_{0} = 2.83\times10^3$
(line 1), $a_{0}=8.0\times10^3$ (line 2) and in CP laser field with $a_{0}=2.0\times10^3$
(line 3), $a_{0}=5.66\times10^3$ (line 4), along with the electric field
strength (in arbitrary units) at the $B$-node location
(dotted line) and the magnetic field strength (in arbitrary units) at 
the $E$-node location (dashed line).}
\label{gamma}
\end{figure}

\begin{figure}
\includegraphics[width=8cm]{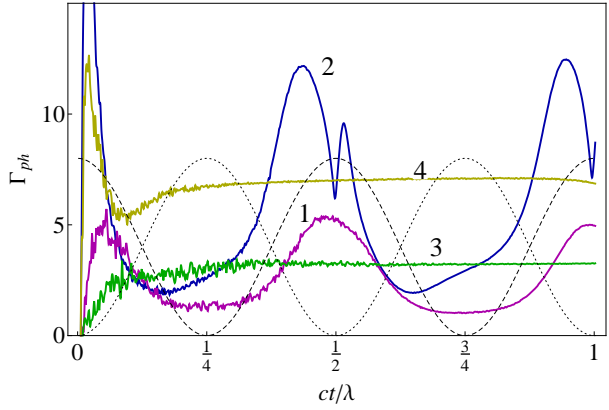}
\caption{The photon number growth rates (solid lines) as function
of $ct/\lambda$ in LP laser field with $a_{0}=2.83\times10^3$
(line 1), $a_{0}=8.0\times10^3$ (line 2) and in CP laser field with $a_{0}=2.0\times10^3$
(line 3), $a_{0}=5.66\times10^3$ (line 4). The electric field
strength (in arbitrary units) at the $B$-node
location (dotted line) as function
of $ct/\lambda$ and the magnetic field strength (in arbitrary units) at 
the $E$-node location
(dashed line) as function
of $ct/\lambda$.}
\label{gammaph}
\end{figure}

\begin{figure}
\includegraphics[width=8cm]{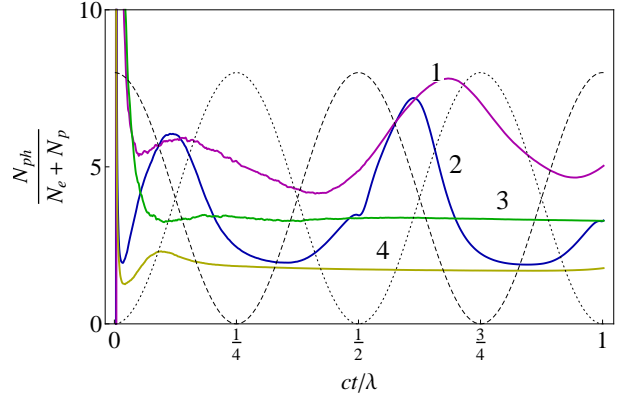}
\caption{Time dependence of the ratio of the numbers of
photons and pairs in LP laser field with $a_{0}=2.83\times10^3$
(line 1), $a_{0}=8.0\times10^3$ (line 2) and in CP laser field with $a_{0}=2.0\times10^3$
(line 3), $a_{0}=5.66\times10^3$ (line 4), along with the electric field
strength (in arbitrary units) at the $B$-node location
(dotted line) and the magnetic field strength (in arbitrary units) at 
the $E$-node location 
 (dashed line).}
\label{ratio}
\end{figure}


\begin{figure}
\includegraphics[width=8cm]{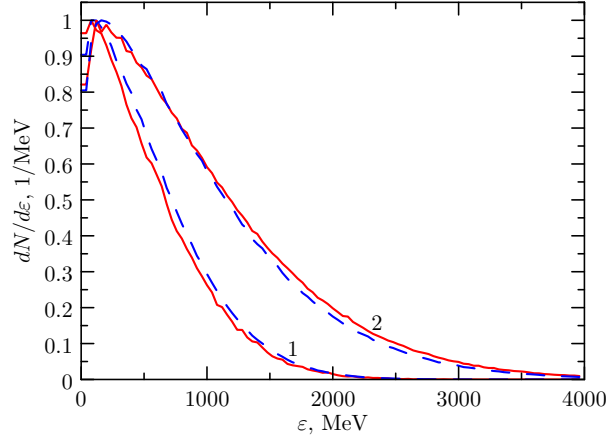}
\caption{The normalized energy spectra of the electron-positrons pairs produced
in QED cascade in CP laser field at the instance $t=1.6\lambda/c$
(red solid line 1) and $t=2\lambda/c$ (blue dashed line 1) for $a_{0}=2.0\times10^3$;
the same at $t=0.6\lambda/c$ (red solid line 2) and $t= 1.0 \lambda/c$ (blue dashed
line 2) for $a_{0}=5.66\times10^3$.}
\label{fig-spectrum-c}
\end{figure}

\begin{figure}[h]
\begin{minipage}[c]{ \linewidth}%
  \center{\includegraphics[width=8cm]{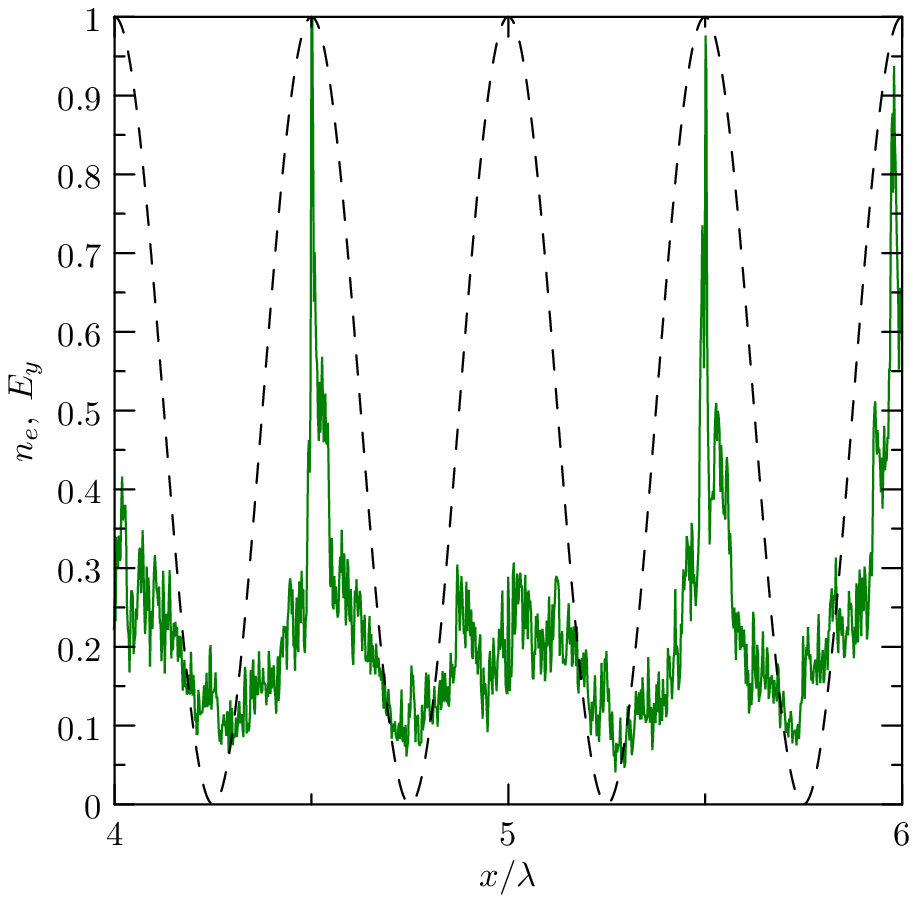}
\\
  a)} %
\end{minipage}\vfill%
\begin{minipage}[c]{ \linewidth}%
  \center{\includegraphics[width=8cm]{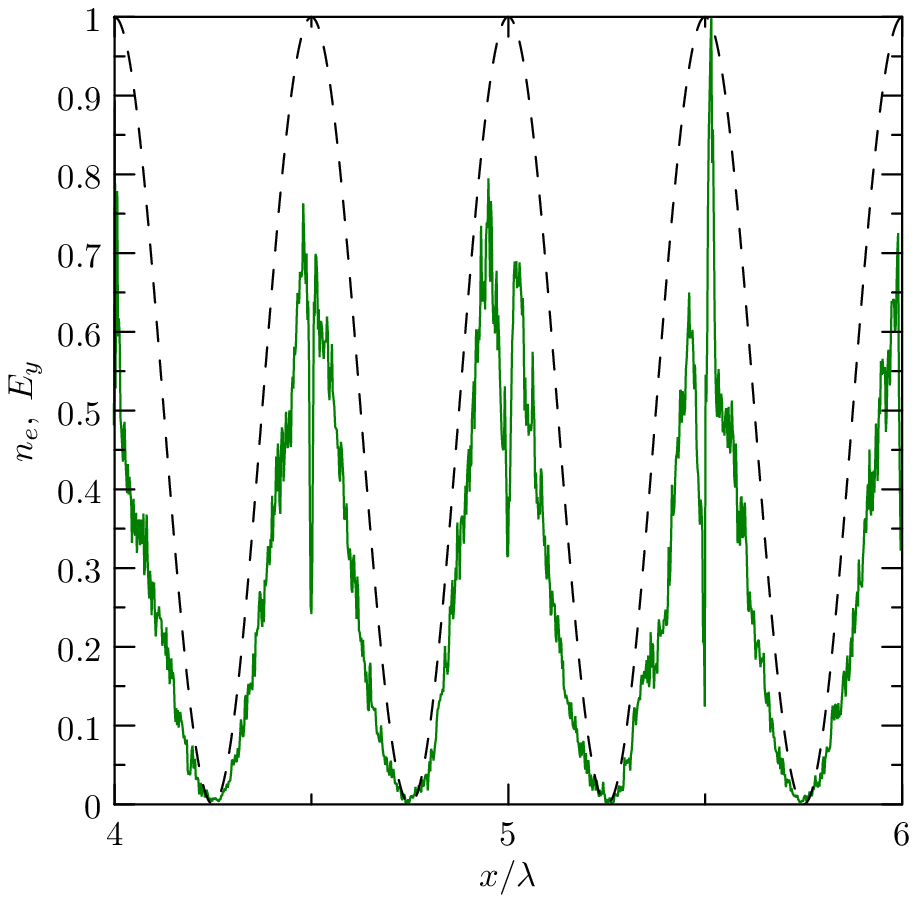}
\\
  b)} %
\end{minipage}\caption{Electric field strength normalized to the  amplitude (black
dashed line) and pair density normalized to it's maximum value
(green solid line) as functions of $x$ at the instance $t=2\lambda/c$ in CP laser
field for a) $a_{0}=2.0 \times 10^3$, b) $a_{0}=2^{-1/2}\cdot 8000$.}
\label{fig-density-c}
\end{figure}

The results of the simulation are shown in Figs.~\ref{fig-N}-\ref{fig-density-c}.
It follows from Fig.~\ref{fig-N} that the number of pairs is growing exponentially. $N\propto\exp(\Gamma t)$, up to $4\times10^7$ and $2\times 10^{13}$ during a 
laser period for $a_{0}=2.0\times 10^{3}$ and $a_{0}=5.66\times 10^3$.
Accordingly, the cascade growth rate can be estimated as $\Gamma\simeq 3.3 \omega_L$ and $\Gamma\simeq 7.1 \omega_L$, respectively (see Figs.~\ref{gamma}
 and \ref{gammaph}).
In particular, the inverse cascade growth rate is much shorter than the
laser period. The ratio of the numbers of photons and pairs is about 
$3.4$ for and $1.7$ (see Fig.~\ref{ratio}). 
As can be observed from Fig.~\ref{fig-density-c}, the electron-positron plasma is
produced mostly near the plane $B=0$. The normalized energy spectra of the electrons
and positrons produced in the cascade are shown in Fig.~\ref{fig-spectrum-c}
at two random successive time instances for both value of $a_0$, it becomes clear that the shape of the spectra remains conserved in time. The mean energy of electrons and positrons is around $500$~MeV 
 for $a_{0}=2.0\times10^3$ and $1$~GeV for 
$a_{0}=5.66\times10^3$.

\subsection{Linear polarization}

Now consider QED cascading driven by linearly polarized (LP) laser
pulses. The components of the laser field at $t=0$ are
\begin{eqnarray}
E_y & = & a_0 g(y,z)\left[f_1(x+x_0)-f_1(x-x_0)\right],\label{elp} \\
E_z & = & B_y=0,\nonumber\\
B_z & = & a_0 g(y,z)\left[f_1(x+x_0)+f_1(x-x_0)\right].\label{blp}
\end{eqnarray}

  In order to simplify the mapping with the previous CP case, we asume the same power and energy of the pulses, thus considering the value $a_0=2^{1/2}\cdot 2000=2.83\cdot10^3$
and $a_0=2^{1/2}\cdot2000=8.0\times10^3$, with all the other parameters being the same as before.
As is shown in Fig.~\ref{fig-N}, the number of pairs is growing during a laser
period this time up to $10^7$ and $2\times10^11$ for the chosen value of $a_0$. However, in the present case the cascade growth rate oscillates from  $\sim0.9\omega_{L}$ to 
$\sim4.5\omega_{L}$
for $a_{0}=2.83\times10^3$ and from $\sim1.8\omega_{L}$ to 
$\sim10.4\omega_{L}$
for $a_{0}=8.0\times10^3$ (see Figs.~\ref{gamma}
 and \ref{gammaph}). The photon-pair ratio oscillates from $4.2$ to $7.8$ 
and from $1.9$ to $7.2$, respectively (see Fig.~\ref{ratio}). Hence,
unlike the CP case, the number of the cascade particles is increasing in
time stair-step-like at the logarithmic scale.

\begin{figure}[h]
\begin{minipage}[c]{ \linewidth}%
  \center{\includegraphics[width=8cm]{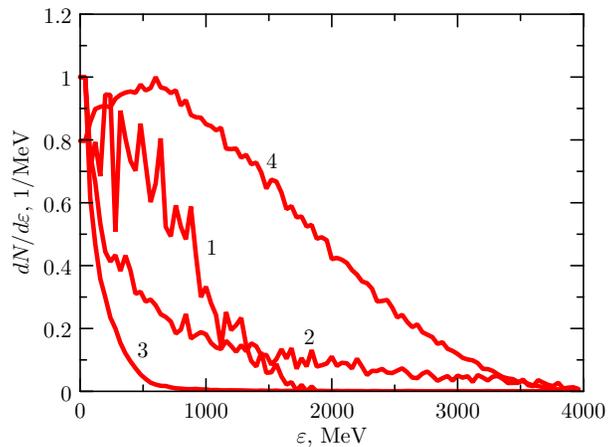}
\\
  a)} %
\end{minipage}\vfill%
\begin{minipage}[c]{ \linewidth}%
  \center{\includegraphics[width=8cm]{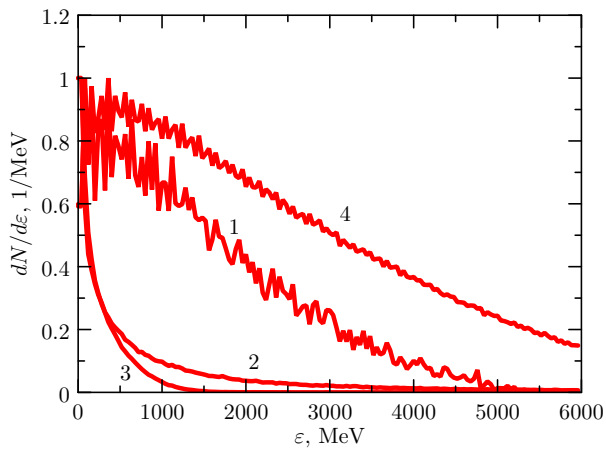}
\\
  b)} %
\end{minipage}\caption{The normalized energy spectra of the 
electron-positron pairs produced
in QED cascade in LP laser field for a) $a_{0}=2.83\times10^3$ at the time instances
  $t=1.2 \lambda/c$ (line 1), $t=1.4 \lambda/c$ (line 2), $t=1.6\lambda/c$
  (line 3), $t=1.8\lambda/c$ (line 4); b) $a_{0}=8.0\times10^3$ at the time instances $t=0.2\lambda/c$
  (line 1), $t=0.4\lambda/c$ (line 2), $t=0.6\lambda/c$ (line 3),
  $t=0.8\lambda/c$ (line 4).}
\label{fig-spectrum-l}
\end{figure}

\begin{figure}
\includegraphics[width=8cm]{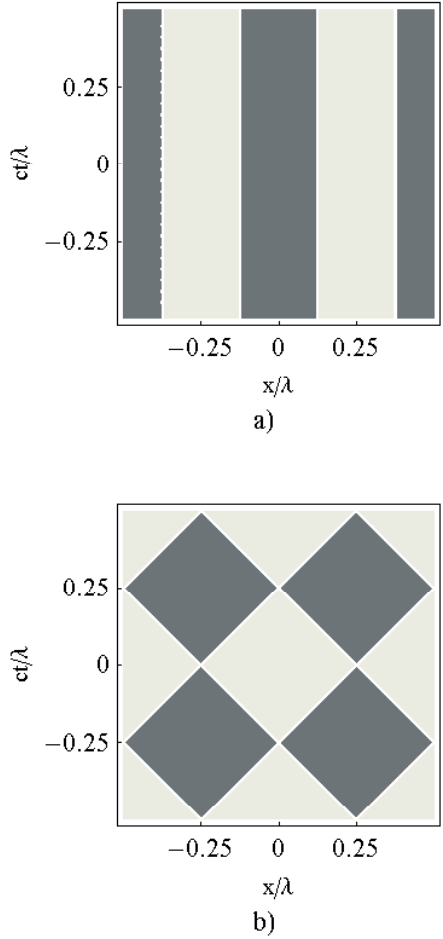}
\caption{The ``electric'' region, where the electric field is stronger 
than the magnetic one, (dark gray area) and 
``magnetic'' region, where the magnetic field is stronger 
than the electric one,  (light gray area) in $x-t$ plane for
the circularly (a) polarized standing electromagnetic 
wave and for the linearly polarized standing electromagnetic 
wave (b).}
\label{region}
\end{figure}

\begin{figure}[h]
\begin{minipage}[c]{ \linewidth}%
  \center{\includegraphics[width=8cm]{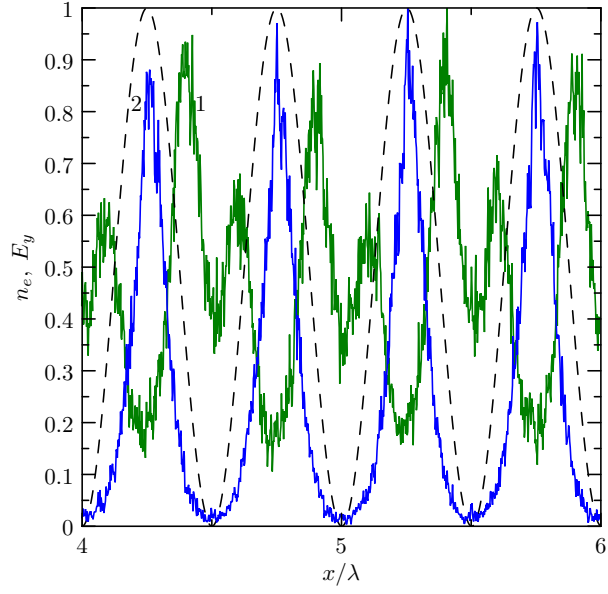}
\\
  a)} %
\end{minipage}\vfill%
\begin{minipage}[c]{ \linewidth}%
  \center{\includegraphics[width=8cm]{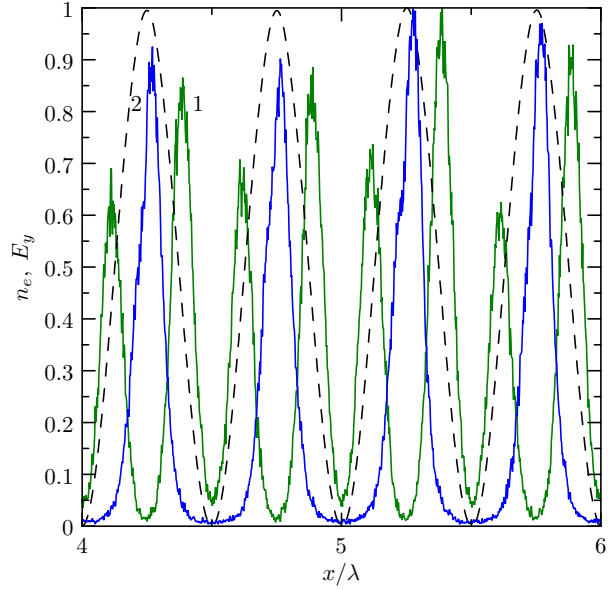}
\\
  b)} %
\end{minipage}\caption{The electric field strength normalized to the 
filed amplitude (red dashed line) and the plasma density normalized to the density maximum
(green solid line) at the time instance $t=1\lambda/c$ in LP laser field 
for a) $a_{0}=2.83\times10^3$, b) $a_{0}=8.0\times10^3$.}
\label{fig-density-l}
\end{figure}

  The energy spectra of the electrons and positrons produced in a QED
cascade is depicted in Fig.~\ref{fig-spectrum-l} for several successive time moments. 
The distribution function of the cascade particles is breathing with the period, which equals to a half of the laser period. During $0.2\lambda/c<t<0.4\lambda/t$ a moderate 
	growth of the number of particles accompanying by plasma heating can be observed. Note that this time interval stands out for the the volume of the spatial region where  
$|\mathbf{E}(x,t)|>|\mathbf{B}(x,t)|$   (the "electric'' region) is larger than of the region where 
$|\mathbf{E}(x,t)|>|\mathbf{B}(x,t)|$   (the "magnetic'' region)
 and that the electron-positron plasma is mostly   located near 
the plane $E=0$ (see Fig.~\ref{region}). During $0.4\lambda/c<t<0.55\lambda/c$  
 particle production  
peaks but the mean energy decreases. Lastly, during $0.55\lambda/c < t < 0.7\lambda/c$ particle production 
becomes strongly suppressed but the mean energy per particle reaches a minimum. For this time interval the volume of the "electric'' region 
  becomes smaller than that of the "magnetic'' region and the electron-positron
  plasma density has two maximums around each of the planes $E=0$ (see 
Fig.~\ref{fig-density-l}).

The mean energy per particle oscillates between a small value and $1.5$~GeV for $a_0=2.83\times10^3$ or $3$~GeV for $a_0=8.0\times10^3$,
respectively (see Fig.~\ref{fig-spectrum-l}). It peaks at $t=0.25\lambda/c$ and $7=0.75\lambda/c$ when the "electric'' region occupies all space and becomes minimal at $t=0.5\lambda/c$ and $t=\lambda/c$ when per contra the "magnetic" region extends to all the space.

\subsection{Comparison of cascading in CP and LP fields}

The time dependence of the number of particle in a cascade can be generally parametrized as $N(t)\propto exp\left(\int\Gamma dt\right)$, 
where $\Gamma$ is the instantaneous cascade growth rate. In the LP case the growth rate $\Gamma_{LP}(t)$ is a periodical function with the period being a half of the laser period, and so o the energy spectra. 
Accordingly, the number of particles is growing stair-step-like at the logarithmic scale. This contrasts the CP case, in which the cascade growth rate $\Gamma_{LP}(t)$ and pair spectra remain constant while the particle number is growing exponentially. 

The number of cascade particles produced at $t=2\lambda/c$ in LP 
field is on the case $a_{0}=2.83\times10^3$ approximately a quarter of those  in CP  field with the corresponding value $a_{0}=2.0\times10^3$. As for $a_0=8.0\times10^3$  the number of
particles produced in CP field is already a hundred times less than in LP field for $a_0=5.66\times10^3$. 
Thus, more particles are produced in CP field than in LP field for the same laser energy.  

Eliminating the oscillations, the ratio of the numbers of particles for linear and circular polarisations can be cast in the form $N_{CP}/N_{LP}\simeq\exp\left[(\Gamma_{CP}-\left\langle \Gamma_{LP}\right\rangle)t \right]$, where $\left\langle \Gamma_{LP}\right\rangle $ is the average over a period of the LP case growth rate. For the two sets of $a_0$ under consideration this quantity is around $1.4\omega_L$ and $4.7\omega_L$, respectively. One can also introduce a polarisation factor as the ratio of the cascade growth rates for LP and CP cases, $\left\langle \Gamma_{LP}\right\rangle/\Gamma_{CP}$, which acquires the value $0.85$ and $0.8$, respectively, i. e. has an advantage that it depends rather weakly on $a_0$. As for the mean energy per particle, for our parameters it's found around three times higher in LP case than in CP case. This is rather natural in a view of the previous discussion, because cascading obviously tends to suppress most of all the high energy population of plasma particles.

\section{Analytical model}
\subsection{General consideration}

In this Section we develop simple analytical model for QED cascading
in the standing electromagnetic wave of arbitrary polarization so
that the field components are functions of $x$ and $t$ only. The
dynamics of the cascade particles is governed by the kinetic equations
\cite{Rossi1953,Nerush2011}. However the cascade kinetic equations
cannot be solved analytically in general case. Here we will use more
simple approach based on analysis of cascade particle dynamics 
\cite{Fedotov2010}. For simplicity we assume that the cascade 
particles double within the time interval much lower than the 
laser period, that is $\Gamma\gg1$. We also assume 
that  $\chi\gg1$  for the most of 
electrons and positrons when they emit photons and $\chi_{ph}\gg1$ 
for the most photons when they decay with electron-positron pair 
production. In the limit $\chi\gg1$ the energy for the most of 
the electrons (positrons) after photon emission is much lower 
than that before emission.

The continuity equation for the cascade particle density can be
written as follows
\begin{equation}
\frac{\partial n}{\partial t}+\frac{\partial}{\partial x}\left(v_{x}n\right)
-\Gamma n=0,
\label{discont}\end{equation}
where $n$ is the density of the cascade particles and $v$ is the
particle velocity. As $\Gamma\gg1$, $v_{x}<1$ and
the particle displacement between QED events is small $\delta x\sim1$
the second term in Eq.\,(\ref{discont}) can be neglected. Therefore 
we can conclude that the most of the particles are produced at a given
space point rather than come from neighbourhood locations and we can
exclude the space motion of the cascade particles from consideration.
Neglecting the motion of the cascade particles, the equations for
the numbers of particles take a form:
\begin{eqnarray}
\frac{dN_{e+p}}{dt} & = & 2W_{pair}N_{\gamma},\label{Nep}\\
\frac{dN_{\gamma}}{dt} & = & W_{em}N_{e+p}-W_{pair}N_{\gamma},
\label{Ngam}\end{eqnarray}
  where $N_{e+p}$ is the number of the electrons and positrons and
$N_{\gamma}$ is the number of the photons. Solving the equations
we find that $N\propto\exp\Gamma t$, where
\begin{eqnarray}
\Gamma & = & \frac{W_{pair}}{2}\left(-1+\sqrt{1+\frac{8W_{rad}}
{W_{pair}}}\right),\label{Gamma1}\\
\frac{N_{\gamma}}{N_{e+p}} & = & 
\frac{\Gamma}{2W_{pair}},\label{NeNg}\end{eqnarray}
  where Eqs.\,(\ref{W3}) and (\ref{Wg3}) are used for $W_{rad}$
and $W_{pair}$, respectively.

To estimate the cascade growth rate we should calculate temporal evolution
of $\gamma$ and $\chi$ for the test cascade particle. The electron
dynamics between the time moments of photon emission is governed by
equations of motion
\begin{eqnarray}
\frac{d\boldsymbol{\mathbf{p}}}{dt} & = & -\boldsymbol{\mathbf{E}}
-\left[\frac{\mathrm{\mathbf{p}}}{\gamma}\times\mathbf{B}\right],
\label{motion1}\\
\frac{d\mathbf{r}}{dt} & = & \frac{\mathrm{\mathbf{p}}}{\gamma}.
\label{motion2}\end{eqnarray}
  where $\boldsymbol{\mathbf{p}}$ is normalized to $mc$, $\gamma$
is gamma-factor of the particle, $t$ is normalized to $\omega_{L}^{-1}$,
the coordinates are normalized to $c/\omega_{L}$, the electromagnetic
field strength is normalized to $mc\omega_{L}/|e|$. The equation for
positron motion can be derived from Eqs.\,(\ref{motion1}) and 
(\ref{motion2})
replacing $e$ by $-e$. In the laser field with normalized field
strength $a$ the gamma-factor of the particle is limited by $a$.
As the electron lost most of its energy after photon emission we suppose that the electron is initially ($t=t_{0}$) at rest (just after photon emission).

The characteristic times of elementary cascade processes
like photon emission and pair production are much smaller than laser
period $t_{rad},\: t_{pair}\ll1$ for typical cascade conditions 
\cite{Fedotov2010,Elkina2011},
where $t_{rad}\approx W_{rad}^{-1}$ and $t_{pair}\approx W_{pair}^{-1}$
are the characteristic times of photon emission and pair production,
respectively. Therefore we can solve Eqs.\,(\ref{motion1}) and 
(\ref{motion2})
expanding solution in Taylor series in $\delta t\ll1$ near $t=t_{0}$.
The first-order term of $\gamma$ can be presented as follows $\gamma=
\left(\delta t\right)ak_{\gamma}(x_{0},t_{0})$,
where $k_{\gamma}(x_{0},t_{0})$ is a function of the electromagnetic
field strength in the initial time instant and in the initial electron
position $x_{0}=x(t=t_{0})$. It follows from Eq.\,(\ref{chi}) that
the parameter $\chi$ is approximately equal to the product of $\gamma$
and the force component which is transverse to the electron momentum.
The last is vanishing at $t=t_{0}$ as the electron first moves along
the force direction. So, $\chi=\left(\delta t\right)^{2}a\eta 
k_{\chi}(x_{0},t_{0})$, as $\chi \propto \phi a \gamma $ and 
the angle between the electron velocity and Lorentz force is 
$\phi\sim\delta t$
to the first order in $\delta t$, where 
$\eta=\hbar\omega_{L}/\left(mc^{2}\right)$
and $k_{\chi}(x_{0},t_{0})$ is again function of the field components
at the initial moment of time and in the initial electron position.
The $k$-factors for CP and LP standing wave are calculated in Appendixes.

Combining formulas for $\gamma$, $\chi$ and $W_{rad}$ the closed
system of equations for the electron (positron) can be derived
\begin{eqnarray}
\gamma & \approx & at_{rad}k_{\gamma},\\
\chi & \approx & a^{2}t_{rad}^{2}\eta k_{\chi},\label{chi0}\\
W_{rad} & \approx t_{rad}^{-1} & 
\approx1.4\alpha\eta^{-1}\gamma^{-1}\chi^{2/3}.
\label{system1}\end{eqnarray}
  $t_{rad}$ can be excluded from the system so that the system can
be expressed through the electromagnetic field parameters
\begin{eqnarray}
\chi & \approx & (a_{*}\eta k_{\gamma})^{3/2},\label{chi1}\\
\gamma & \approx & a_{*}^{3/4}\eta^{1/4}k_{\chi}^{-1/2}k_{\gamma}^{7/4},
\label{gamma1}\\
W_{rad} & \approx & 1.4\alpha a_{*}^{1/4}\eta^{-1/4}k_{\chi}^{1/2}
k_{\gamma}^{-3/4},
\label{system2}\end{eqnarray}
  where $a_{*}=a(1.4\alpha)^{-1}$. As the photon absorbs substantial 
portion of
the electron energy and it is emitted in the direction of the electron velocity just before emission, we can assume for
the sake of simplicity
$\gamma_{ph}=\gamma\gg1$ and $\chi_{ph}=\chi\gg1$ so that
$W_{rad}\approx(1.46/0.38)W_{pair}$
\begin{eqnarray}
\Gamma & \approx & 1.22W_{rad},
\label{rate}\\
\frac{N_{\gamma}}{N_{e+p}} & \approx & 2.34.
\label{NgNe}\end{eqnarray}
It's worth to note that this relations are universal and valid for both 
circular polarisation and "electric" region of linear polarisation for 
the high intensities. We can find the numerical confirmation of this 
assertion in the Fig.\,\ref{ratio}, where the line~4 
(circular polarisation, high intensity) and parts of line~2 
(linear polarisation, high intensities), corresponding to the 
electric region, is in agreement with  Eq.\,(\ref{NgNe}).
Making use of Eq.\,(\ref{chi1}), we can estimate $\chi\approx2.14$
for $a_{0}=2^{1/2}\cdot2000$ and $\chi\approx10.18$ for $a_{0}=8000$,
where $k_{\chi}$ and $k_{\gamma}$ are assumed to be of the order
of unity, $a=2a_{0}$ is taken into account for the standing wave
and $\lambda=0.91\mu m$. Therefore, the model better fits for numerical
simulations with higher $a_{0}$.

The model presented above is not valid for QED cascading in LP standing
plane wave in the ``magnetic'' space-time region where 
$|\mathbf{B}|>|\mathbf{E}|$.
As $\mathbf{B}\perp\mathbf{E}$ for LP plane wave we can choose the
reference frame where $\mathbf{E'}=0$ at the given time
moment at the given position. It is shown in Appendix that the electron 
dynamics
in the ``magnetic'' region is close to the superposition of
electron Larmor rotation and the slow drift without significant energy
gain. Moreover, the photon emission leads to a rapid electron cooling
there.

In the "magnetic'' frame the field can be considered as static 
and homogeneous as $t_{rad}\ll1$ and the theory 
developed by Akhiezer \textit{et. al} \cite{Akhiezer}
for QED cascading in a magnetic field can be used to analyse cascade
dynamics. In this case the energy of
the cascade particles is limited by the energy of the first particle which
initiated the cascade. The theory predicts that the cascading and
particle production occur until the time moment when the energy of
the cascade particles will be so low that $\chi$ becomes lower than
$1$ for all particles. The estimates for the total number of the
produced particles, $N_{B}$, and the characteristic time of cascade
development, $t_{B}$, can be obtained from the theory:
\begin{eqnarray}
N_{B} & = & \gamma a\eta,\label{Nb}\\
t_{B} & = & \frac{81}{32}\left[\Gamma\left(\frac{4}{3}\right)
\Gamma\left(\frac{2}{3}\right)\right]^{-1}\frac{\gamma^{1/3}}{q}\gamma_{B},
\label{tb}\\
V_{b} & = & \frac{|\mathbf{E}|}{|\mathbf{B}|},\label{vb}\\
\gamma_{B} & = & \frac{|\mathbf{B}|}{\sqrt{\mathbf{B}^{2}-\mathbf{E}^{2}}},
\label{gb}
\end{eqnarray}
  where $V_{b}$ and $\gamma_{B}$ are the velocity and the gamma-factor
determining the "magnetic'' reference frame, respectively, $t_b$ is a 
characteristic cascade duration (which can be estimated as a number of 
events $\log\chi_0$ times time of a one event $1/W$) and
$q$ is given by
\begin{equation}
q=\frac{\alpha3^{1/6}}{2\pi}\Gamma\left(\frac{2}{3}\right)\left(\frac{a^{2}}
{\eta^2 (1+V_{B})^{2}\gamma_{B}^{2}}\right)^{1/3}.
\label{q}\end{equation}
 The derived equations can be applied to estimation of the cascade
growth rate for linear polarization in ``magnetic'' region
\begin{equation}
\Gamma_{B}\approx\frac{1}{t_{B}}\ln N_{B}.
\label{Gb}\end{equation}
Although the particle number increase non-exponentially 
in the ``magnetic'' region we have introduced $\Gamma$ by the same way
as it had been done for the exponential growth in the ``electric'' region
(see Eq.~(\ref{Gamma1})).

\subsection{Circular polarization}

First we analyse CP as the most simple type of polarization. The
dimensionless vector-potential, electric and magnetic fields of such
field configuration are given by
\noindent \begin{eqnarray}
\mathbf{A} & = & a(0,\cos x\sin t,\cos x\cos t),\label{A_c}\\
\mathbf{E} & = & a(0,\cos x\cos t,-\cos x\sin t),\label{E_c}\\
\mathbf{B} & = & a(0,-\sin x\cos t,\sin x\sin t).\label{B_c}
\end{eqnarray}
  where fields rotate around the $x$-direction. The invariant
  $\mathcal{F}=\mathbf{E}^{2}-\mathbf{B}^{2}$
for LP standing wave takes a form
\begin{equation}
\mathcal{F}=a^{2}\cos2x.
\label{Fcp}
\end{equation}
  It follows from Eq.\,(\ref{Fcp}) that $\mathcal{F}$ is conserved
in time for CP standing wave. As It follows from the definition of $\mathcal{F}$
(see also Fig.~\ref{ratio})
that $\mathcal{F} > 0$ in the ``electric'' region,  $\mathcal{F} < 0$ in 
the ``magnetic'' region and $\mathcal{F} = 0$ on the border between regions. 
The coefficients $k_{\chi}$ and $k_{\gamma}$
for CP are derived explicitly in Appendix:
\begin{equation}
k_{\chi}(x)=\sqrt{\frac{\cos^{2}x}{\tan^{2}x+4}},\ k_{\gamma}=\cos x_{0}.
\label{k-circ}\end{equation}
  The factors are time-independent as well as $\mathcal{F}$. At $x=0$
we have $k_{\chi}=1/2$, $k_{\gamma}=1,$ and Eqs.\,(\ref{system2})
reduced to that derived in Ref.\,\cite{Fedotov2010}. Making use
of Eqs.\,(\ref{system2}) and (\ref{k-circ}) the cascade growth rate
can be calculated. Analysis of the rate shows that the cascade rate
is almost constant for $-\pi/3+\pi l<x<\pi/3+\pi l,\ l\in Z$. This is 
close to
  what follows from plasma distribution obtained in numerical
  simulation for high-intensity example (see Fig.\,\ref{fig-density-c} b)).

The model ratio of the photon number to the pair number is given by
Eq.\,(\ref{chi1}) and is close to the value obtained from numerical
simulation for $a_{0}=2^{-1/2}\cdot8000$ (see Fig.\,\ref{ratio}).
The model predicts that $\Gamma\approx5.6$, $\epsilon\approx400$\,MeV
for $a_{0}=2000$ and $\Gamma\approx6.8$ for $a_{0}=2^{-1/2}\cdot8000$,
$\epsilon\approx800$\,MeV. As expected the prediction for
$a_{0}=2^{-1/2}\cdot8000$
is in better agreement with numerical results demonstrated in
Figs.\,\ref{fig-spectrum-c}
and Figs.\,\ref{gamma} than that for the low-intensity case.

\subsection{Linear polarization}

Now we analyse QED cascading in LP standing wave. The dimensionless
vector-potential, electric and magnetic fields are

\noindent \begin{eqnarray}
\mathbf{A} & = & a(0,-\cos x\sin t,0)\label{A_l}\\
\mathbf{E} & = & a(0,\cos x\cos t,0),\label{E_l}\\
\mathbf{B} & = & a(0,0,\sin x\sin t).\label{B_l}
\end{eqnarray}
Please note that the phase of the laser field given by Eqs.~(\ref{elp})
and (\ref{blp}) is shifted by $\pi /2 $ from the phase of the electromagnetic
wave given by Eqs.~(\ref{A_l})-(\ref{B_l}.) 
  The normalized QED parameter $\mathcal{F}$ for LP standing wave
takes a form
\begin{equation}
\mathcal{F}(x,t)=\cos2t+\cos2x,\label{Flp}
\end{equation}
  where the parameter is normalized to $a^{2}/2$. It follows from
Eq.\,(\ref{Flp}) that $\mathcal{F}$ is a periodic function of
time with the half of the laser period and the volume of the ``electric''
and ``magnetic'' regions evolves in time (see Fig.~\ref{region}). The ``electric''
region occupies all space twice per laser period at $t=\pi l$,
$l\in Z$, while the ``magnetic'' region expands up to all space
at $t=\pi/2+\pi l$, $l\in Z$. Some electrons and positrons produced
in the cascade can be first accelerated in the electric region and
then radiate their energy in the magnetic region. Even immobile particle
can be in the ``electric'' region at some time moments and in
the ``magnetic'' region at the other time moments because the
boundary between ``electric'' and ``magnetic'' regions oscillates.
Therefore, the cascade dynamics in LP field is more complex than
that in CP field.

\noindent In the ``electric'' region we can use Eq.\,(\ref{system2}),
where coefficients $k_{\chi}$ and $k_{\gamma}$ are calculated in
Appendix:
\begin{eqnarray}
k_{\chi}^{2}(x,t) & = & \frac{\mathcal{F}(x,t)\tan^{2}x
\left(\cos^{2}x+\sin^{2}t\right)^{2}}{8\cos^{2}x\cos^{2}t},
\label{kchilp}\\
k_{\gamma}^{2}(x,t) & = & \mathcal{F}(x,t).
\label{kglp}
\end{eqnarray}
  As the coefficients depend on time, the cascade growth rate is also a
function of time which agrees with Fig.\,(\ref{rate}). In general
the contribution to the cascade growth rate is given by both ``electric''
and ``magnetic'' regions simultaneously. To compare our model
with numerical results we consider time moments $t=\pi l$, $l\in Z$,
when the ``electric'' region occupies all space. We introduce
the cascade growth rate averaged over $x$ in the ``electric''
region as follows
\begin{eqnarray}
N(t) & = & N(t_{0})\exp\left(\int_{t_{0}}^{t}\bar{\Gamma}_{E}(t')dt'\right),
\label{naver}\\
\bar{\Gamma}_{E}(t) & = & \frac{\int n(x,t)\Gamma(t,x)dx}{\int n(x,t)dx}
\label{gaver}
\end{eqnarray}
  where by the definition
  \begin{equation}
n(x,t)=n(x,t_{0})\exp\left(\int_{t_{0}}^{t}\Gamma(t',x)dt'\right).
\label{nlp1}\end{equation}
Making use of the electron distribution shown in Fig.\,\ref{fig-density-l}
and Eqs.\,(\ref{rate}), (\ref{kchilp}), (\ref{kglp}) we can estimate
$\bar{\Gamma}_{E}\approx1.95$ for $a_{0}=2^{1/2}\cdot2000$ and
$\bar{\Gamma}_{E}\approx2.53$
for $a_{0}=8000$, which is in a fairly good agreement with numerical results
for $ct=0.25\lambda$ and $ct=0.75\lambda$, respectively (see 
Fig.\,\ref{gamma}).
The particle density peaks near $x=0$ for the time moments $t=\pi l$,
$l\in Z$. However $\chi\approx0$ at $x=0$ as follows from 
Eqs.\,(\ref{kchilp}),
(\ref{chi0}), because the charged particle moves strictly along the
electric field \cite{Bulanov2010} as always $B=0$ at $x=0$. The
particles are produced around point $x=0$ in the region where $\chi>1$
and reaches $x=0$ because this point is attractive for the electrons
and positrons during half of the laser period. At the position where
$\chi\approx1$ near $x=0$ we can estimate $k_{\chi}\approx1$ and
$k_{\gamma}\approx2$ so that the mean particle energy is 
$\epsilon\approx1500$\,MeV
for $a_{0}=2^{1/2}\cdot2000$ and $\epsilon\approx3000$\,MeV for
$a_{0}=8000$ that is in a good agreement with the numerical results
(see Fig.\,\ref{fig-spectrum-l}).

Now let us analyse cascading in the ``magnetic'' region with 
$|\mathbf{B}|>|\mathbf{E}|$. It is shown in 
Appendix that the particle acceleration
is suppressed in this region and the electrons and positrons lose
almost all their energy because of photon emission. Cascading and
particle production occur until the energies of
the cascade particles is so low that $\chi<1$ for all particles.
Let us assume that the electron gains the energy in the ``electric''
region. Then the boundary between ``electric'' and ``magnetic''
regions is shifted so that the electron finds oneself in the ``magnetic''
region. We can estimate the growth rate of the cascade initiated by
the electron using Eqs.\,(\ref{Nb}), (\ref{tb}). The result is
$\Gamma_{B}\approx5.6$ for $a_{0}=2^{1/2}\cdot2000$ and $\Gamma_{B}\approx14.5$
for $a_{0}=8000$. We can conclude that the particle production is
more efficient in the ``magnetic'' regions than in the ``electric''
ones, which is in qualitative agreement with the numerical results
(see Fig.\,\ref{gamma}) demonstrating the enhanced particle
production when the ``magnetic'' region dominates. 

There are two reasons why particles are produced more efficiently  
in the ``magnetic'' region. Firstly, In the ``electric'' region electrons and
positrons are accelerated by the laser field so the angle between
the particle momentum and the Lorentz force is small. In the ``magnetic'' region
the particles are not accelerated and the angle can be large
thereby increasing $\chi$ and enhancing the probability
of the particle production. Secondly, the particle energy decreases 
in time in the ``magnetic'' region because of photon emission. 
This also enhances the particle production probability as the probability
increases with decreasing of the particle energy 
(see Eqs.~(\ref{W3}) and (\ref{Wg3})).

The quantitative comparison of the cascade growth rate predicted 
by the model with that obtained numerically is difficult because 
cascades develops in both ``electric'' and magnetic'' regions 
permanently (see Fig.~\ref{region}). The ``magnetic'' region 
occupies all space at the time moments $t=\pi/2+\pi l$, $l\in Z$. 
However the electrons and positrons are strongly cooled by these time
moments so that the cascading is suppressed and $\Gamma \simeq 1$. 
Therefore the self-consistent theory including QED cascading instantaneously 
in both ``electric'' and ``magnetic'' regions is needed.

\section{Conclusions}

In Conclusion we study QED cascading in the field of two counter-propagating
laser pulses for both circular and linear polarizations.
We restricted ourself by initial stages of the cascade  when the particle 
number is small so that the self-generated plasma fields do not affect
cascade dynamics. First the cascade dynamics is explored by numerical simulation
with 3D PIC-MC code. The particle number increase mostly exponentially in time.
The cascade growth rate, the particle spectra and the distribution of the 
produced plasma tend to be constant in time for CP laser field while they 
periodically evolve with half of the laser period for LP laser field.
It is shown that for a given laser energy the number of the particles 
produced in the cascade with the CP laser field is greater than in 
the LP one. 

We develop simple analytical model of QED cascading in the standing
plane electromagnetic wave. The model is based on the analysis of the single particle dynamics. 
For simplicity we consider the limit $\Gamma \gg 1$.
In this limit most of the particles are produced at a given
space point rather than come from neighbourhood locations and we can
exclude the space motion of the cascade particles from consideration.
However even for low intensity case $a_0 < 3000$ when the parameter $\chi$
is of the order of the unity the model is in a qualitative agreement with
the numerical results.

The model can explain some key features of the cascade. 
The cascade dynamics is governed by relativistic invariant $\mathcal{F}$.
In the CP standing wave $\mathcal{F}$ is constant in time, 
and the particle spectra and cascade growth rate become being stationary.
In contrast, in the LP standing wave $\mathcal{F}$
oscillates in time with half of the laser period, which leads to 
the stair-step-like dependence of the particle number 
on time and periodical evolution of the particle spectra.
For LP laser field the cascade dynamics in the "electric" region 
(where electric field is stronger than magnetic one) 
is strongly dissimilar from that in the "magnetic" region (where the magnetic field 
is stronger than electric one). In the "electric" region the electrons and 
positrons can be accelerated by the laser field up to very high energy.
Unlike that the lepton acceleration is suppressed in the "magnetic" region. 
Moreover, the high-energy leptons are cooled by photon emission.
The spectrum evolution predicted by the model is in good agreement 
with the results of numerical simulation. 
As the volume of the "electric" and "magnetic" regions evolves periodically in 
time the cascade growth rate for LP laser field is also 
a periodic function of time with the period equal to the half of the 
laser period. 

The model estimation of the cascade growth rate for circular polarization  
is in a good agreement with the numerical result even for low-intensity 
example when the model assumption $\chi \gg 1$ is not strictly fulfilled.
The quantitative comparison of the cascade growth rate predicted by the 
model for linear polarization with that obtained numerically is difficult 
because the cascading occurs simultaneously in both ``electric'' and 
``magnetic'' regions most of time while  $\Gamma $ can be calculated 
only if cascade develops only in one of two regions.  The self-consistent
theory for linear polarization  including QED cascading instantaneously 
in both ``electric'' and ``magnetic'' regions is needed.
To explain the dynamics of the self-generated plasma distribution the 
self-consistent theory should also include the temporal dynamics of the plasma 
density and should be extended to the particles with $\chi < 1$ as the number 
of such particles is large especially in the low-intensity example.  

\begin{acknowledgements}
This work was partially supported in part by the Government of the Russian 
Federation (Project No. 14.B25.31.0008), by the Russian Foundation for 
Basic Research (Grants No 13-02-00886, 13-02-00372), by Russian Federation President's grant (Grant No МК-5853.2013.2),
by the Federal Targeted Programme “Scientific and Scientific-Pedagogical Personnel of the Innovative Russia in 2009-2013” 
(Governmental Contract No. 14.A18.21.0773), and by the President Grants for Government Support of the Leading Scientific Schools 
of the Russian Federation (grant No. NSh-5992.2012.2).
\end{acknowledgements}

\appendix

\section{Electron dynamics in CP standing wave}

Let's introduce normalized parameter $\chi^{2}_0$:
\begin{equation}
\chi^{2}_0 = a^{-2}\eta^{-2}\chi ^{2},
\label{a1chi1}\end{equation}
where $\chi $ is given by Eq.~(\ref{chi}), $\mathbf{p}$ is the momentum 
normalized
to $mc$ and the field
strength is normalized to $a\omega_{L}mc/e$. In the CP case electric
and magnetic fields are parallel to each other: $\mathbf{B}=s\mathbf{E}$,
where $s=-\tan x_{0}$. Taking into account that, we derive for 
$\chi_{0}^{2}$
\begin{equation}
\chi_{0}^{2}=\left(s^{2}+1\right)\left[\left(p_{z}E_{y}-p_{y}E_{z}\right)^{2}
+p_{x}^{2}\mathbf{E}^{2}\right]+\mathbf{E}^{2}.
\label{a1chi2}\end{equation}
  The last term can be neglected as the electron is accelerated to
the relativistic energy within very short time period $t_{rel}\sim 
a^{-1}\ll t_{rad},\: t_{pair}\ll1$..

We suppose that the electron was at rest at $x=x_{0}$ and $t=t_{0}$.
We choose axis $y$ along 
$\mathbf{E_{0}}=\mathbf{E}(x=x_{0},t=t_{0})=s\mathbf{B}(x=x_{0},t=t_{0})$,
so that $E_{z}=B_{z}=0$ at $x=x_{0}$ and $t=t_{0}$. We can expand
the fields near $x=x_{0}$ and $t=t_{0}$\begin{eqnarray}
E_{y} & \approx & E_{0}+\left(\partial_{t}E_{y}\right)\delta t
+\left(\partial_{x}E_{y}\right)\delta x,\label{a1ey}\\
E_{z} & \approx & \left(\partial_{t}E_{z}\right)\delta t
+\left(\partial_{x}E_{z}\right)\delta x,
\label{a1ez}\end{eqnarray}
  where $\delta t\ll1$ and $\delta x\ll1$. We can also expand the
electron momentum components:\begin{eqnarray}
p_{x} & \approx & \frac{1}{2}p_{x}''\delta t^{2},\label{a1px}\\
p_{y} & \approx & \gamma\approx p_{y}'\delta t,\label{a1py}\\
p_{z} & \approx & \frac{1}{2}p_{z}''\delta t^{2},\label{a1pz}
\end{eqnarray}
  where it is taken into account that the electron first moves along
$y$-axis so that $p_{x},\: p_{z}\ll p_{y}$ and $p_{x}'\approx 
p_{z}'\approx0$.

Making use of equation of motion - Eq.\,(\ref{motion1}) we can derive
\begin{eqnarray}
p_{x}''\delta t & \approx & \frac{\left(p_{y}'\delta t\right)
\left(s\delta t\partial_{t}E_{z}\right)-\left(\frac{1}{2}p_{z}''
\delta t^{2}\right)sE_{0}}{p_{y}'\delta t},
\label{a1px2}\\
p_{y} & \approx & \gamma\approx E_{0}\delta t,\label{a1py2}\\
p_{z}''\delta t & \approx & \left(\partial_{t}E_{z}\right)\delta t
+\frac{\left(\frac{1}{2}p''_{x}\delta t^{2}\right)sE_{0}}{p'_{y}\delta t},
\label{a1pz2}
\end{eqnarray}
  where $\mathbf{B}=s\mathbf{E}$ is used and the terms, which are
proportional to $\delta x$, are neglected as $v_{x}=p_{x}/\gamma\ll1$
and $\delta x\ll\delta t$. The solution of Eqs.\,(\ref{a1px3})-(\ref{a1pz3})
is
\begin{eqnarray}
p_{x} & \approx & \frac{s}{4+s^{2}}\left(\partial_{t}E_{z}\right)\delta 
t^{2},\label{a1px3}\\
p_{y} & \approx & \gamma\approx E_{0}\delta t,\label{a1py3}\\
p_{z} & \approx & 
\frac{2(2+s^{2})}{4+s^{2}}\left(\partial_{t}E_{z}\right)\delta t^{2}.
\label{a1pz3}\end{eqnarray}
  Combining Eq.\,(\ref{a1chi2}), Eqs.\,(\ref{a3vb}), (\ref{a3gb})
and Eqs.\,(\ref{a1px3})-(\ref{a1pz3}) $\chi_{0}^{2}$ can be derived
\begin{eqnarray}
\chi_{0}^{2} & \approx & \left( s^{2}+1 \right) \delta t^{4} \left[ d^2_1
+\frac{1}{4}\left(p_{x}''\right)^{2}E_{0}\right]
\label{a1chi3} \\
d_1 & = & \frac{1}{2}p_{z}''E_{0} - p'_{y}\left(\partial_{t}E_{z}\right)
\end{eqnarray}

Finally we get
\begin{eqnarray}
\chi^{2} &=& 2a^{4}\eta^{2}\delta t^{4}k_{\chi}^{2},\label{a1chi4}
\\
k_{\chi}^{2}&=&\frac{\cos^{2}x_{0}}{\tan^{2}x_{0}+4},
\label{a1kchi}
\end{eqnarray}
  where Eqs.\,(\ref{a1px3})-(\ref{a1pz3}) for CP field distribution
is used. The equation for gamma-factor of the electron can be written
as follows
\begin{equation}
\frac{d\gamma}{dt}=a\left(\mathbf{p}\cdot\mathbf{E}\right).\label{a1gt}\end{equation}
  Finally the gamma-factor is
\begin{equation}
\gamma=ak_{\gamma}\delta t,\: 
k_{\gamma}^{2}=\cos^{2}x_{0}.\label{a1g}\end{equation}

\section{Electron dynamics in LP standing wave in the ``electric'' region
($|\mathbf{E}|>|\mathbf{B}|$)}

First we consider the space-time region where $|\mathbf{E}|>|\mathbf{B}|$.
It is convenient to treat the problem in another reference frame,
namely in the ``electric'' frame where at $t=t_{0}$ 
$\mathbf{B}'(x'_{0},t'_{0})=0$
(accent marks quantities in the ``electric'' reference frame).
The appropriate boost's velocity is given by

\begin{equation}
V_{E}=\frac{B_{z}(x_{0},t_{0})}{E_{y}(x_{0},t_{0})},\label{a2v0}\end{equation}
  where $B_{z}(x_{0},t_{0})$ and $E_{y}(x_{0},t_{0})$ are the electric
and magnetic fields in the laboratory reference. Nearby ($x'_{0},$
$t'_{0}$) the field component can be expanded up to the first order:
\begin{eqnarray}
E'_{y} & \approx & E'_{0}+\left(\partial_{t'}E'_{y}\right)\delta t'
+\left(\partial_{x'}E'_{y}\right)\delta x',
\label{a2ez}\\
B'_{y} & \approx & \left(\partial_{t'}B'_{z}\right)\delta t'+\left
(\partial_{x'}B'_{z}\right)\delta x',
\label{a2by}\end{eqnarray}
  It should be noted that for the field derivatives we can write
  \begin{eqnarray}
\left(\partial_{t'}E'_{y}\right) & = & -\left(\partial_{x'}B'_{z}\right),
\label{a2ezby}\\
\left(\partial_{x'}E'_{y}\right) & = & -\left(\partial_{t'}B'_{z}\right)',
\label{a2byez}\end{eqnarray}
  Equation of the electron motion (\ref{motion1}) can be solved with
expansion in time series $a^{-1}\ll\delta t'\ll1$: \begin{eqnarray}
p'_{y} & \approx & -E'_{0}\delta t'-\frac{\left(\delta t'\right)^{2}}
{2}\left(\partial_{t'}E'_{y}\right),\label{a2pz}\\
p'_{x} & \approx & \frac{\left(\delta t'\right)^{2}}{2}
\left(\partial_{x'}E'_{y}\right),
\label{a2px}\end{eqnarray}
  where the terms of the zeroth order on $1/a$ are kept and the terms,
which are proportional to $\delta x$ are neglected because
$\delta x\sim a^{-1}\ll\delta t'$.
The leading term for $\chi$ takes a form
\begin{equation}
\chi=\frac{1}{2}E'_{0}\left(\partial_{t'}B'_{z}\right)
\left(\delta t'\right)^{2}.
\label{a2chi1}\end{equation}
  $\chi$ is the relativistic invariant. Expressing it in terms of
the laboratory-frame quantities and substituting the field component
for LP standing wave from Eqs.\,(\ref{gaver}), (\ref{B_l}) we
obtain \begin{eqnarray}
\chi & \approx & \delta t^{2}a^{2}k_{\chi},\label{a2chi2}\\
k_{\chi}^{2} & = & \frac{\mathcal{F}(x_{0},t_{0})\tan^{2}x_{0}
\left(\cos^{2}x_{0}+\sin^{2}t_{0}\right)^{2}}{8\cos^{2}x_{0}
\cos^{2}t_{0}},
\label{a2kchi}\end{eqnarray}
  where $\mathcal{F}(x_{0},t_{0})$ is the normalized QED parameter
defined by Eq.\,(\ref{nlp1}). It follows from Eqs.\,(\ref{a1px3})
that $\gamma'\approx\left|p'_{y}\right|\approx E'_{0}\delta t'$.
Expressing it in terms of the laboratory-frame quantities and substituting
the field component for LP standing wave we obtain \begin{eqnarray}
\gamma & \approx & ak_{\gamma}\delta t,\label{a2g}\\
k_{\gamma} & = & \mathcal{F}(x_{0},t_{0}).\label{a2kg}\end{eqnarray}

\section{Electron dynamics in LP standing wave in the ``magnetic'' region
($|\mathbf{B}|>|\mathbf{E}|$)}

Let us now consider the space-time region where $|\mathbf{B}|>|\mathbf{E}|$.
It is again convenient to treat the problem in another reference frame,
namely in the ``magnetic'' frame where at $t=t_{0}$ 
$\mathbf{E}'(x'_{0},t'_{0})=0$
(accent marks quantities in the ``magnetic'' reference frame).
The appropriate boost velocity and the boost gamma-factor are given
by
\begin{eqnarray}
V_{B} & = & \frac{E_{y}(x_{0},t_{0})}{B_{z}(x_{0},t_{0})},\label{a3vb}\\
\gamma_{B} & = & 
\frac{B_{z}(x_{0},t_{0})}{\sqrt{B_{z}^{2}(x_{0},t_{0})-E_{y}^{2}(x_{0},t_{0})}}.
\label{a3gb}\end{eqnarray}
  where $B_{z}(x_{0},t_{0})$ and $E_{y}(x_{0},t_{0})$ are the electric
and magnetic fields in the laboratory reference. For simplicity we
will consider region, where $B_{y}>2^{1/2}E_{z}$ so that $\gamma_{B}\sim1$
and $V_{B}\gamma_{B}<1$. Nearby ($x'_{0}$, $t'_{0}$) the field
component can be expanded up to the first order:
\begin{eqnarray}
E'_{y} & \approx & \left(\partial_{t'}E'_{y}\right)\delta t'+
\left(\partial_{x'}E'_{y}\right)\delta x',\label{a3ez}\\
B'_{z} & \approx & B'_{0}+\left(\partial_{t'}B'_{z}\right)\delta t'
+\left(\partial_{x'}B'_{z}\right)\delta x',
\label{a3by}\end{eqnarray}
  The field derivatives obey Eqs.\,(\ref{a2ezby}), (\ref{a2byez}).
We suppose that in the laboratory reference frame the electron is
at rest at the initial moment of time $t=t_{0}$ so that in the magnetic
frame $\gamma'_{0}=\gamma_{B}$, $p'_{x,0}=-v_{B}\gamma_{B}$, 
$p'_{z,0}=p'_{y,0}=0$.
Assuming again that $\delta t'\ll1$ and keeping the leading terms,
the equation of the electron motion (\ref{motion1}) can be rewritten
in the non-relativistic limit as follows

\begin{eqnarray}
\left(\partial_{t'}p'_{x}\right) & = & -B'_{0}p'_{z},\label{a3eq1}\\
\left(\partial_{t'}p'_{y}\right) & = & -t'\left(\partial_{t'}
E'_{y}\right)-B'_{0}p'_{x}.
\label{a3eq2}\end{eqnarray}
  The derived equations describe Larmor rotation of the electron in
the magnetic field with growing electric field. The solution takes
a form:\begin{eqnarray}
p'_{x} & = & p'_{x,0}\cos\left(B'_{0}\delta t'\right)-\delta t'
\frac{\left(\partial_{t'}E'_{y}\right)}{B'_{0}},\label{a3pz}\\
p'_{y} & = & -p'_{x,0}\sin\left(B'_{0}\delta t'\right)
+\frac{\left(\partial_{t'}E'_{y}\right)}{\left(B'_{0}\right)^{2}}.
\label{a3px}\end{eqnarray}
  The terms proportional to $\delta x$' in the field expansion are
neglected Eqs.\,(\ref{a3eq1}), (\ref{a3eq2}) because it follows
from Eqs.\,(\ref{a3pz}), (\ref{a3px}) that $\delta x'\sim
p'_{x,0}/a+\left(\delta t'\right)^{2}/2\ll\delta t'$,
where estimation $B'_{0}\sim\left(\partial_{t'}E'_{y}\right)\sim a$
is used.

Making use of Eqs.\,(\ref{a3pz}), (\ref{a3px}) and the inverse
Lorentz transformation the electron energy gain can be derived in
the laboratory frame
\begin{eqnarray}
\gamma & = & \gamma_{B} \sqrt{\gamma_{B}^{2}+2 d_2 d_3 + d_3^2}
- p_B (d_2 - d_3) ,
  \label{a3dg1}
\\
d_2 &=& p_{B} \cos\left(B'_{0}\delta t'\right) ,
\\
d_3 &=& \delta t' \frac{\partial_{t'}E'_{y}}{B'_{0}} .
  \end{eqnarray}
  Expressing it in terms of the laboratory-frame quantities and substituting
the field component for LP standing wave, we obtain
\begin{eqnarray}
\gamma & = & \gamma_{B}\sqrt{\gamma_{B}^{2}+2k_{B}p_{B}\delta
t\cos\left(\omega_{B}\delta t\right)+\delta t^{2}k_{B}^{2}}\\
  & - & p_{B}^{2}\cos(\omega_{B}\delta t) + k_{B}p_{B}\delta t,
  \label{a3dg2}\\
\gamma_{B} & = & \frac{2^{1/2}\sin x_{0}\sin 
t_{0}}{\sqrt{-\mathcal{F}(x_{0},t_{0})}},
\label{a3gb1}\\
p_{B} & = & \frac{2^{1/2}\cos x_{0}\cos 
t_{0}}{\sqrt{-\mathcal{F}(x_{0},t_{0})}},
\label{a3pb1}\\
\omega_{B} & = & 
\frac{B_{0}}{\gamma_{B}^{2}}=a\frac{-\mathcal{F}(x_{0},t_{0})}
{2\sin x_{0}\sin t_{0}},
\label{a3wb}\\
k_{B} & = & \frac{\partial_{t'}E'_{y}}{\gamma_{B}B'_{0}}=\frac{\sin2x_{0}}
{-\mathcal{F}(x_{0},t_{0})}.
\label{a3kb}\end{eqnarray}
  where $B_{0}=B_{z}\left(x_{0},t_{0}\right)$. It follows from 
Eq.\,(\ref{a3dg2})
that $\gamma\sim1$ and thus there is no significant electron acceleration
in contrast to the ``electric'' region where $\gamma\sim a\delta t\gg1$
(see Eq.\,(\ref{a2g})).

\end{document}